\documentstyle[preprint,aps,prl,epsfig]{revtex}
\tightenlines

\newcounter{listfig}

\begin{document}

%for 2 columns
%\twocolumn[\hsize\textwidth\columnwidth
%\hsize\csname@twocolumnfalse\endcsname
%\draft
%-----------

\title{The quantum dynamics of atomic magnets, co-tunneling and dipolar-biased tunneling}

\author{R. Giraud$^{1,*}$, A.M. Tkachuk$^2$, B. Barbara$^1$}

\address{$^1$ Laboratoire de Magn\'etisme Louis N\'eel, CNRS, BP166, 38042
Grenoble Cedex-09, France\\
$^2$ All-Russia Scientific Center ``S.I. Vavilov State Optical Institute'',
199034 St. Petersburg, Russia\\
$^*$ Present address:  Laboratoire de Spectrom\'etrie Physique, BP 87, 38402 Saint Martin d'H\`eres, France
}

\date{\today}

\maketitle

\begin{abstract}

Multi-spins tunneling cross-relaxations in an ensemble of weakly-coupled Ho$^{3+}$ 
ions, mediated by weak anisotropic dipolar interactions, can be evidenced by ac-susceptibility 
measurements in a high temperature regime. 
Based on a four-body representation, including the rare-earth nuclear spin, two-ions tunneling mechanisms can be attributed to 
both dipolar-biased tunneling and co-tunneling processes. The co-reversal involving entangled pairs of magnetic moments is 
discussed with a particular emphasis, giving new evidences to elucidate the many-body quantum dynamics.

%PACS numbers: 75.45.+j, 71.70.-d, 75.50.Lk
\end{abstract}
\pacs{75.45.+j, 71.70.-d, 75.50.Lk}

%for 2 columns
%]
%-----------

\narrowtext

Since the discovery of quantum tunneling of the magnetization in single molecule magnets \cite{Thomas96,BB95,Friedman96}, 
the effects of environmental degrees of freedom on tunneling of mesoscopic spins, 
especially due to other spins \cite{Prokofev00}, have been investigated in detail within the single spin'
framework \cite{Tupitsyn00,Giraud01}. Indeed, most features of the staircase-like hysteresis loops, observed in an \emph{ensemble} of weakly-interacting mesoscopic spins, were rather well 
understood within this simple picture of a single spin submitted to external and internal magnetic fields. In the presence of an axial 
anisotropy, hysteresis occurs at low temperatures due to the energy barrier $E_B$ hindering the magnetization reversal, when thermal activation 
over the barrier requires a very long relaxation time $\tau = \tau_0 e^{E_B/k_BT}$. Nevertheless, 
non-stationnary quantum mechanics allows for a faster time-evolution at avoided level crossings within the single-spin Zeeman diagram (tunneling), leading to magnetization steps for well-defined 
values of the magnetic field applied along the easy axis (resonance conditions). Each mesoscopic spin experiences a time-dependent distribution of 
internal fields which is determinant for the analysis of the quantum relaxation of weakly-coupled spins \cite{Prokofev00,Thomas99,Ohm99}. These fields being generated by environmental spins, 
coupled by dipolar and eventually exchange interactions, one should ask the question to what extent the measured tunneling rate is related to 
quasi-isolated spins or to larger entities such as two (or more) spin centres.  

Recent magnetization measurements of mesoscopic Ho$^{3+}$ ions, highly diluted in a LiYF$_4$ single crystal, evidenced the coherent rotation of both the electronic 
angular momentum and the nuclear spin of a nearly-isolated Ho$^{3+}$ ion \cite{Giraud01}. However, some additional small magnetization steps were also observed 
at fast field-sweep rates, and still remain ambiguous because these steps have no explanation in term of level crossings 
within the single-ion model, including the rare-earth nuclear spin. In this letter, we demonstrate that new tunneling mechanisms originate from an \emph{unused} feature of this simple picture: 
equally-spaced energy levels allow for non-dissipative spin--spin cross-relaxation tunneling processes. Ac-susceptibility measurements in a high 
temperature regime give us a unique mean to assess the \emph{various} multi-spins tunneling mechanisms which may affect the quantum dynamics of an assembly of 
mesoscopic spins. In particular, we focus on the tunneling of pairs of Ho$^{3+}$ ions, weakly coupled by anisotropic dipolar interactions, leading to new resonance conditions that 
are an intrinsic property of the two-ions picture (no matter is the coupling strength). We call this phenomenon a '\emph{co-tunneling}' process, and show that 
pairs of Ho$^{3+}$ ions may co-tunnel with interactions as weak as a few mK. This suggests that, under these specific field conditions, any 
environmental Ho$^{3+}$ ion may co-tunnel with a '`central'' spin, making the quantum spin-dynamics much more collective than expected before. 

At low temperatures, a nearly isolated Ho$^{3+}$ ion in LiYF$_4$ behaves as an \emph{atomic magnet} due to a large axial crystal-field anisotropy along the 
tetragonal easy $c$-axis \cite{Giraud01}, thus leading to very long spin--phonon relaxation times. Therefore, the magnetic 
moment is oriented along the $c$-axis and constitutes a two-level system (electronic doublet of Ising-type), with the moment being frozen either 'up' or 'down'. However, 
the latter may still evolve due to quantum tunneling at avoided electro-nuclear level crossings of a single Ho$^{3+}$ ion, the rare-earth nuclear 
spin being $I=7/2$. The single-ion tunneling process thus results from the coherent rotation of both its electronic 
momentum and nuclear spin, the crossing fields being only related to the hyperfine constant $A_J$ \cite{Giraud01}. In this '\emph{strong} hyperfine coupling case, 
decoherence only occurs due to nuclear spins of diamagnetic ions (that is, mainly the nuclear spins of eight fluorine ions, with $I=1/2$). 
Furthermore, the effect of a many-body quantum dynamics was also observed in holmium-based spin glasses \cite{Aeppli91}, but the way how quantum fluctuations 
arise in this more concentrated regime still remains unclear. 

\begin{figure}
%preprint
\centerline{\epsfxsize= 15 cm \epsfbox{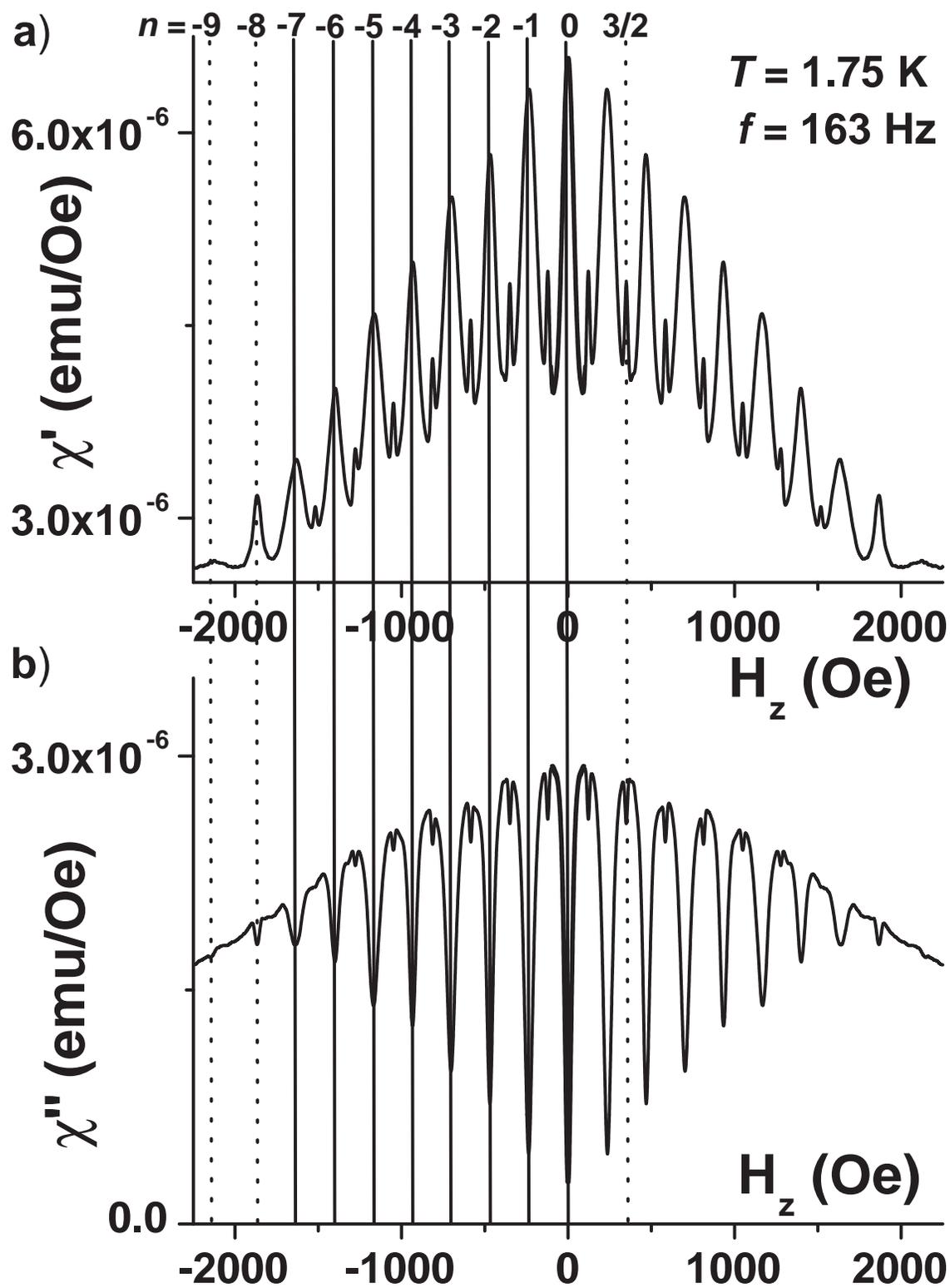}}
\caption{R. Giraud}
\label{fig1}
\end{figure}

\begin{figure}
%preprint
\centerline{\epsfxsize= 15 cm \epsfbox{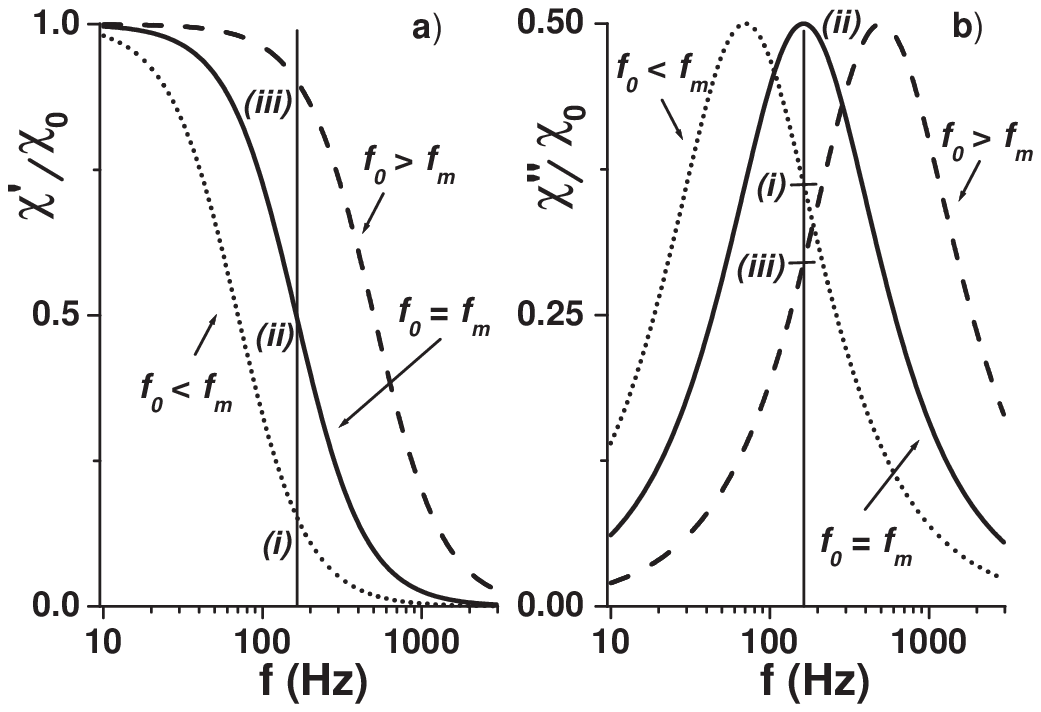}}
\caption{R. Giraud}
\label{fig2}
\end{figure}

Ac-susceptibility of a 0.1$\%$ at. holmium-doped LiYF$_4$ single crystal was studied as a function of a quasi-static field applied along the easy $c$-axis. 
The measurements were performed in a high temperature regime ($T\geq 1.75$~K) with a Quantum Design MPMS SQUID magnetometer, using a 4~Oe-amplitude excitation field. 
In Fig.~\ref{fig1}, a low-frequency measurement, as obtained with $f_m = 163$~Hz, reveals peaks (dips) in the in-phase (dissipative) response when the dc-field is changed, allowing us to investigate the role of the various quantum relaxation mechanisms. In every cases, the results can be qualitatively 
understood when considering a field-dependent relaxation time to describe the system dynamics. As shown in Fig.~\ref{fig2}, if the constant 
frequency measurement $f_m$ is smaller than the characteristic frequency $f_0$ of spin fluctuations (case -'\emph{iii}-), an increase in the in-phase response, or a decrease in the 
dissipative response, is associated with a faster relaxation rate (that is, $f_0$ increases). The large deviations to the monotonous response, as observed in Fig.~\ref{fig1}, are clear 
signatures of the \emph{single-ion quantum dynamics} in a thermally-activated tunneling process through the energy barrier ($E_B\approx 10$~K, due to the first 
excited electronic singlet). Indeed, the faster relaxation of the magnetization observed for all the resonant fields $H_n$ (see solid lines in Fig.~\ref{fig1}), 
with $H_n = n\times 230$~Oe ($-7\le n \le +7$, $n$ being an integer), is in agreement with the crossing-fields of the electro-nuclear Zeeman diagram calculated 
within the single-ion picture (shown in Fig.~\ref{fig4}a). Moreover, their width is temperature-dependent. Besides, similarly to non-expected magnetization steps 
observed in hysteresis loops at very low temperatures, additional peaks/dips are also evidenced for integer values of $\vert n\vert$ larger than 7, as 
well as for half-integer $n$ values with $-13 \le 2n \le +13$, which do not correspond to any level crossing in Fig.~\ref{fig4}a). These strongly suggest that 
\emph{two} new tunneling mechanisms occur. Because of the nature of the intra-atomic hyperfine coupling, all these fields 
involve a unique property within the single-ion representation: some energy levels are equally separated, so that simultaneous multi-spins transitions may 
conserve the total energy. Furthermore, part of these cross-spins relaxation (CSR) processes may \emph{not} conserve the total magnetization of 
the system, some electro-nuclear states being associated with opposite directions of the magnetic moment, thus allowing for a \emph{multi-ions quantum dynamics}. Among these transitions, we can 
clearly distinguish between the \emph{co-tunneling} and the \emph{biased tunneling} processes. In particular, we discuss the role of co-tunneling transitions for 
the first time in mesoscopic magnetism. As shown below, ac-susceptibily anomalies for half-integer $n$ are only related to co-tunneling events involving 
a two-ions process, which only occur within the dipolar window, their width thus being temperature-independent. 
In addition, this width is smaller than the one related to thermally-activated single-ion processes, because at $T \geq 1.75$~K the homogeneous 
broadening due to spin--phonon couplings is larger than the inhomogeneous dipolar broadening. Note that thermal activation over the energy 
barrier still requires longer times than the CSR \emph{tunneling} events; the spin--phonon relaxation time can be mesured for an applied field located 
out of any resonance \cite{Giraud}. In this strongly diluted limit, $N$-ions co-tunneling transitions with $N>2$, allowed at fields $H_p=p\times H_{(n=1)}/N$ 
(integer $p$, with $\vert p \vert \leq 7N$), are less probable. If not, $N = 3$ transitions should be observed, the first resonance 
condition $H_{(n=1)}$/3 being out of the dipolar width of the zero-field resonance. 

\begin{figure}
%preprint
\centerline{\epsfxsize= 15 cm \epsfbox{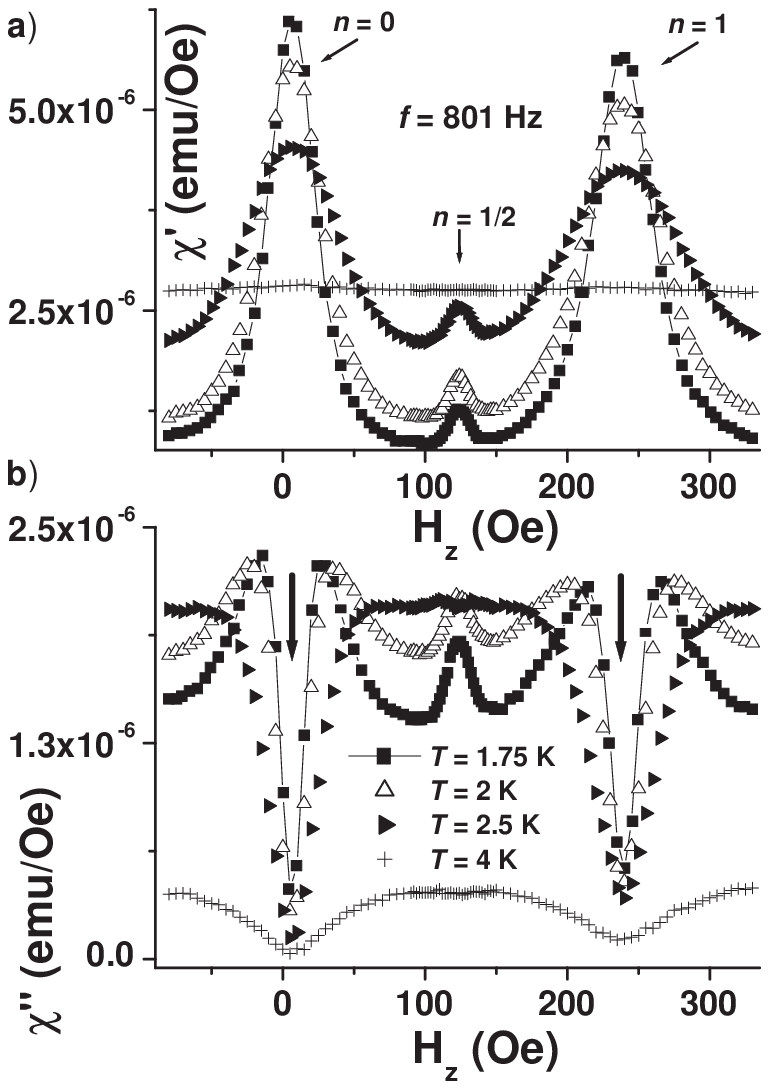}}
\caption{R. Giraud}
\label{fig3}
\end{figure}

\begin{figure}
%preprint
\centerline{\epsfxsize= 15 cm \epsfbox{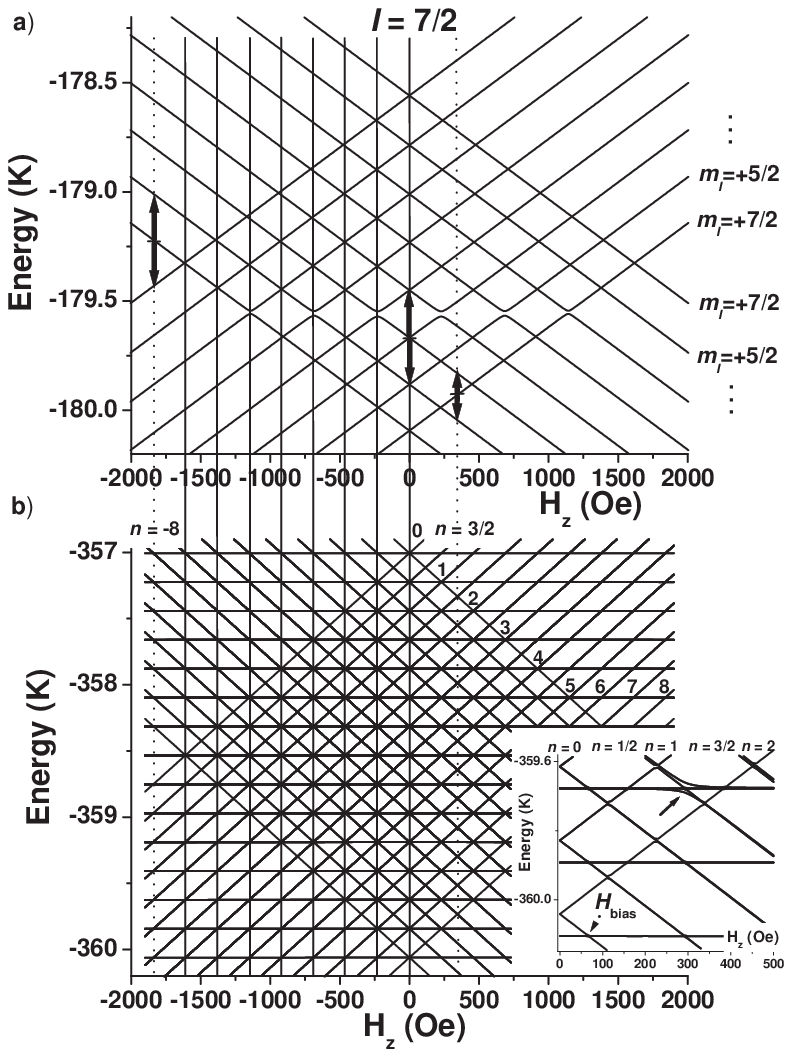}}
\caption{R. Giraud}
\label{fig4}
\end{figure}

Finally, to clearly distinguish between every 
tunneling mechanisms, we used a higher measurement frequency $f_m$ to evidence the role of dipolar-biased tunneling processes. Indeed, these lead 
to additional anomalies in the dissipative response. At $f_m = 800$~Hz and $T = 1.75 $~K, the relaxation rate associated with a thermally-activated 
single-ion tunneling process is smaller than $f_m$ (case -'\emph{i}- in Fig.~\ref{fig2}), whereas it is larger for the dipolar-biased tunneling process (case -'\emph{iii}-). 
Both rates increase being closer to resonance conditions related to integer $n$ values with $\vert n \vert \le 7$. However, the dissipative response now increases close to a 
resonance in the former case, whereas it decreases in the latter one. As the phonon-induced broadening is larger than the dipolar-induced one 
in this temperature range, the response first \emph{increases} when the applied field gets closer to the resonance, but still being out of the dipolar 
window so that CSR do not occur. When this field enters the dipolar window centred around the resonance, the relaxation is 
quickly dominated by the CSR rate and the dissipative response now \emph{decreases}. This feature allows us to clearly evidence, for the first time, 
the dipolar-biased tunneling mechanism as the occurence of a dip dug in the dissipative response, as shown in Fig.~\ref{fig3}b). Again, the width of 
this dip corresponds to the narrow dipolar window, but it is now temperature-independent only below $T \approx 2.2$~K. When the temperature is increased 
further, the thermaly-activated tunneling rate indeed gets faster than the measurement frequency. The large maximum envelope of the response 
then turns into a minimum and we cannot distinguish between these two processes anymore.

To underline the role of spin--spin interactions and discuss the physics of CSR in LiYF$_4$:Ho$^{3+}$ in detail, we go beyond the single-ion picture and 
introduce a two-ions representation energy scheme. As shown in Fig.~\ref{fig4}b), within this four-body picture (two electronic angular momenta and two 
nuclear spins) the former situation of equally-spaced levels now corresponds to the existence of energy level crossings of pair states, 
associated either with a co-tunneling event (that is, two spins, initially in the same state, flip simultaneously to the opposite state), 
a biased tunneling process (that is, only one spin flips) or a flip-flop process (that is, two spins with an opposite initial state flip 
simultaneously). Only the first two mechanisms contribute to the relaxation of the magnetization, the latter being associated with the 
so-called 'spin-diffusion' phenomenon. The two-ions Zeeman diagram was calculated on the basis of the single-ion electro-nuclear states shown 
in Fig.~\ref{fig4}a). Details of the calculation will be published elsewhere. Without taking any interaction between Ho$^{3+}$ ions into account, we underline 
in Fig.~\ref{fig4}b) the intrinsic new features of this four-body picture: some additional crossing-fields are only related to the energy-level 
separations within the single-ion picture (see dotted lines). The effect of an additional \emph{isotropic} coupling is shown in the inset of Fig.~\ref{fig4}b), using a 
large amplitude for the sake of clarity.  
Part of the degeneracies is removed so that, in an ensemble of weakly-coupled mesoscopic spins, some crossings are 
shifted by a distribution of equivalent'molecular fields $H_{bias}$, leading to biased tunneling. Co-tunneling crossing-fields are not affected. Note 
that only \emph{anisotropic interactions} allow for CSR tunneling transitions, leading to finite tunneling gaps, such as the non-secular part of the dipole--dipole interaction. All the 
features observed in Fig.~\ref{fig1} now have an interpretation in terms of level crossings involving a pair of Ho$^{3+}$ ions, that is, within the 
four-body picture. The same ideas can be extended to weakly-coupled single molecule magnets (SMMs), such as the well-known molecular nanomagnets Mn$_{12}$-ac and Fe$_8$. Indeed, 
co-tunneling could explain specific heat anomalies observed at intermediate temperatures in Fe$_8$ by Gaudin and co-workers \cite{Gaudin02}. Moreover, 
isotropic exchanged-biased tunneling was also recently observed in a molecular dimer [Mn$_4$]$_2$ but in a strong antiferromagnetic coupling regime \cite{Wernsdorfer02a}, 
thus preventing co-tunneling events, as well as in weakly-interacting Mn$_4$ SMMs \cite{Wernsdorfer02b}, still with a larger ferromagnetic exchange coupling than dipolar interactions.

In LiYF$_4$:Ho$^{3+}$, two-ions energy levels are strongly degenerate due to the high degree of symmetry of the single-ion Zeeman diagram, the strong 
dipolar hyperfine coupling resulting in many equally-separated energy levels. This unique property makes diluted Ho$^{3+}$ ions 
in LiYF$_4$ a \emph{model system} to evidence, and describe in detail, all the various CSR tunneling mechanisms that can occur in an ensemble of weakly-coupled mesoscopic spins. 
Generally speaking, resonance conditions associated with the multi-spins tunneling dynamics can be easily deduced from energy-conserved 
multi-transitions events within the single-spin picture. In particular, co-tunneling processes are a pure \emph{intrinsic property of 
the many-body picture}, even for vanishing interactions, the crossing fields being independent of the coupling strength. More importantly, co-tunneling events are only allowed 
by anisotropic interactions, which are also expected to be a key ingredient to understand quantum fluctuations at larger scales in complexe systems, such as in the quantum spin glass 
problem.

\newpage

\centerline{LIST OF FIGURES CAPTIONS}

\begin{list}{FIG.~\arabic{listfig}}{\usecounter{listfig}}

\item (a,b) Ac-susceptibility measured at $T = 1.75$~K with a frequency $f_m = 163$~Hz. 
The non-monotonous ac-response, as a function of $H_z$, gives evidence for a faster relaxation at well-defined values $H_n = n\times 230$~Oe. Large and broad peaks 
or dips are clearly observed for integer $n$ values, with $-7 \leq n \leq +7$, whereas smaller and narrower ones are measured for half-integer $n$ values,
 with $-13 \leq 2n \leq +13$, as well as for $n=\pm8$ and $n=\pm9$. The solid lines correspond to electro-nuclear level crossings calculated within the single-ion representation (see Fig.~\ref{fig4}a), whereas 
the dotted lines are associated with level crossings calculated within the two-ions representation (see Fig.~\ref{fig4}b).

\item Variation of the in-phase (a) and dissipative (b) responses in the ac-susceptibility for three different relaxation rates $f_0$, 
described within the Debye model assuming a single (dominant) relaxation time. The constant measurement frequency $f_m$ is suggested by the vertical solid line, 
and $\chi_0$ is the static susceptibility. Depending on the value of $f_m$, the dissipative reponse can be either increased (case -'\emph{i}-) or decreased (case -'\emph{iii}-) 
if the dynamics gets faster. In any case, the in-phase response follows the variations of $f_0$.

\item (a,b) Ac-susceptibility measured at a larger frequency $f_m = 800$~Hz. Dipolar-biased tunneling is clearly evidenced, at low-enough temperatures, due 
to a decrease in the dissipative response for every integer $n$ values. The large arrows in (b) show the dips observed for $n = $~0 and $n = $~1. The 
width of these dips is constant below $T \approx 2.2$~K and corresponds to the average amplitude of dipolar interactions.

\item (a) Electro-nuclear energy levels calculated within the single-ion representation. Equally-spaced levels in the Zeeman diagram allow for non-dissipative 
CSR processes in the tunneling regime. As suggested by red arrows, \emph{three} various kinds of CSR may occur at well-defined values 
$H_n = n\times 230$~Oe. (b) Within the two-ions Zeeman diagram, each cross-relaxation process involving a pair of magnetic moments now corresponds to a 
level crossing. Note that the finite slopes are twice larger than in (a). Their crossing describes a co-tunneling process (integer or half-integer $n$, 
with $-14 \leq  2n \leq  +14$). Other crossings only involve the flipping of one magnetic moment (integer $n$, with $-14 \leq  n \leq  +14$). The inset 
shows how an isotropic interaction, with a given coupling stength, removes part of the degeneracies for integer $n$ values, leading to biased tunneling, as well as an 
avoided level crossing due to both Ho$^{3+}$ nuclear spin and first electronic singlet.

\end{list}

\newpage

\end{document}